# Persistent Homology for Breast Tumor Classification using Mammogram Scans

Aras Asaad[1*], Dashti Ali[2], Taban Majeed[3], Rasber Rashid[3]

[1]School of Computing, The University of Buckingham, UK.
[2] Independent Researcher, Ontario, Canada.
[3]Department of Computer Science, Salahaddin University, Kurdistan Region, Iraq.
aras.asaad@buckingham.ac.uk,
dashti.a.ali@gmail.com, rasber.rashid@su.edu.krd, taban.majeed@su.edu.krd
[*]Corresponding Author.

**Abstract.** An Important tool in the field topological data analysis is persistent Homology (PH) which is used to encode abstract representations of the homology of data at different resolutions in the form of persistence barcode (PB). Normally, one will obtain one PB from a digital image when using a sublevel-set filtration method. In this work, we build more than one PB representation of a single image based on a landmark selection method, known as local binary patterns (LBP), that encode different types of local texture from a digital image. Starting from the top-left corner of any 3-by-3 patch selected from an input image, LBP process starts by subtracting the central pixel value from its eight neighboring pixel values. Then assign each cell with 1 if the subtraction outcome is positive, 0 otherwise, to obtain an 8-bit binary representation. This process will identify a set of landmark pixels to represent 0-simplices and use Vietoris-Rips filtration to obtain its corresponding PB. Using LBP, we can construct up to 56 PBs from a single image if we restrict to only using the binary codes which have 2 circular transitions between 1 and 0. The information within these 56 PBs contain detailed local and global topological and geometrical information which can be used to design effective machine learning models. We used four different PB vectorizations namely, persistence landscapes, persistence images, Betti Curves (Barcode Binning) and PB statistics. We tested the effectiveness of proposed landmark-based PH on two publicly available breast abnormality detection datasets using mammogram scans. Sensitivity and specificity of landmark-based PH obtained is over 90% and 85%, respectively, in both datasets for the detection of abnormal breast scans. Finally, experimental results give new insights on using different PB vectorizations with sublevel set filtrations and landmark-based Vietoris-Rips filtration from digital mammogram scans.

**Keywords:** Topological Data Analysis, Persistent Homology, Breast Mammogram, Persistence Diagram vectorization, Medical imaging, Local Binary Patterns.

# 1 Introduction

Topological Data Analysis (TDA) is a collection of methods from algebraic topology and geometry to build and extract topological features from data. Persistent Homology (PH), the main tool of TDA, extracts topological summaries from data in the form of connected components, loops and cavities using a process known as filtration which relies on a nested sequence of simplicial complexes that capture birth and death of those topological invariants [1]. A collection of births and deaths of those topological features are then represented as points in persistence diagram(s) (PD) or equivalently as bars in persistence barcode(s) (PB). Topological structures represented as PDs are stable with respect to small perturbations to the input data when bottleneck or Wasserstein distance is used to compare PDs [2]. Although mostly used when the input data have the form of a point cloud, PH can also be used when the input data to TDA pipeline are images where they have a grid structure. We demonstrate that one can construct Vietoris-Rips filtration from digital images based on pixel landmark locations that convey different types of local textural information. In this paper, we aim at harnessing the power of PH to differentiate benign breast tumors from their malignant counterparts using breast mammogram. Mammogram scan is a special type of X-ray imaging involves exposing breast tissues with a small amount of radiation to obtain an inside picture of breast details for the purpose of abnormality/mass detection and classification.

Female breast cancer is among the 4 leading types of cancer in women worldwide. The World Health Organisation (WHO) ,and its cancer research agencies such as American Cancer society and International agency for cancer research, reported 19.3 million new cases of cancer in 2020 with 10 million deaths and estimated that this number could be increased to 28.4 million new cases by 2040 [3]. Mammogram scan has a number of advantages to detect early signs of breast cancer in women among them its wide deployment in hospitals, easy to store, less time to examine by radiologists and low cost. A number of difficulties face radiologists to properly examine mammograms such as low resolution, size of the lesion within breast tissue, location of the lesion and dense breast tissue in young patients. Therefore, designing sophisticated computer aided diagnostics (CAD) to assist radiologists in making their final decision is a necessity.

The main contribution of this paper can be summarized as follows: (1) constructing 56 persistence diagrams from a single mammogram whereby each PD constructed based on a set of automatically extracted mammogram pixel locations that convey different type of textural information. (2) The space of persistence barcodes featurised using 4 different methods namely binning, barcode-statistics, persistence images and persistence landscapes to measure the true performance of proposed approach.

# 2 Methods

To build PH from digital mammograms, we rely on pixel-based landmarks that correspond to abnormality in textures. We derive our approach from a texture descriptor method known as local binary patterns (LBP) introduced more than two decades ago in [4]. Abnormality is expected to distort local texture and structure in mammogram scans. Using LBP, we encode this change in local texture and structure of mammograms to ensemble a set of point clouds as input to PH pipeline. This method provides a rich source of persistence topological feature for machine learning. Next, we describe our proposed landmark selection procedure and PH construction.

## 2.1 Image Patch Local Binary Patterns (IP-LBP)

Since 1996 LBP has been used successfully in many pattern recognition applications and different versions of LBP proposed and investigated with considerable success [5][6][7]. For any grayscale image $I$, LBP constructs a new grayscale image $\bar{I}$ by encoding each pixel $p \in I$ with 8-bit binary representation determined by comparing central pixel with that of its eight neighbors in a 3-by-3 image-patch surrounding it in a clockwise manner. Starting from the top-left corner of any 3-by-3 patch, LBP process starts by subtracting the central pixel value from its eight neighboring pixel values. Then assign each cell with 1 if the subtraction outcome is positive, 0 otherwise. See **Fig. 1** for illustration. This process results in an 8-bit binary code which can then be converted back to decimal values representing the central pixel $(x_c, y_c)$ using the following equation:

$$LBP(x_c, y_c) = \sum_{i=1}^{i=8} f(p_i - p_c) \times 2^i \qquad (1)$$

where $p_c$ is the central pixel value, $p_i$ is the neighboring gray-value pixels and the function $f(x)$ is defined as follow:

$$f(x) = \begin{cases} 1 \ if \ x \geq 0 \\ 0 \ if \ x < 0 \end{cases} \qquad (2)$$

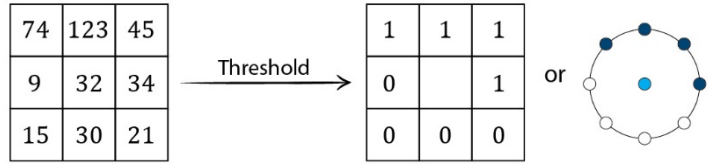

**Fig. 1.** LBP process where 1s in the binary code are represented by bolded points on the circle.

In total, there are 256 binary codes one will obtains for any 3-by-3 image patch following LBP procedure. In [8], Ojala et al. demonstrated that only 58 binary codes, out of the 256, are enough to represent 90% of textures in natural images. The 58 binary codes are known as Uniform LBP (ULBP) and they experimentally demonstrated that the histogram of ULBP codes can be used as a discriminating feature for computer vision applications [9][7]. ULBP codes encode local texture features like edges, corners, spots, lines and flat regions in an image and their binary codes have either 0 or 2 circular transitions from 0 to 1 or from 1 to 0. There are 56 ULBP codes which have 2 circular transitions and only 2 ULBP codes with 0 circular transitions in their 8-bit binary representation. 00000000 and 11111111 are the two ULBP codes with 0 transitions. Examples of ULBP codes with 2 circular transitions are 11000000 and 00111100 whereas a binary code like 10101010 is not a ULBP because there are more than 2 circular transitions from 1 to 0 or vice versa. We can group the 56 ULBP codes according to the number of 1's in their binary representation to form a 7-group geometry $G_\lambda$ for $\lambda = 1, 2, \ldots, 7$ where $\lambda$ refers to the number of 1's in each geometry. Furthermore, each $G_\lambda$ consists of 8 binary codes that can be obtained from each other by a circular rotation, see **Fig 2**. Starting from top-left corner of mammograms, we scan the entire input image by selecting central pixel value of 3-by-3 patches as landmarks if its binary representation satisfies one of the geometrical circles in **Fig 2**.

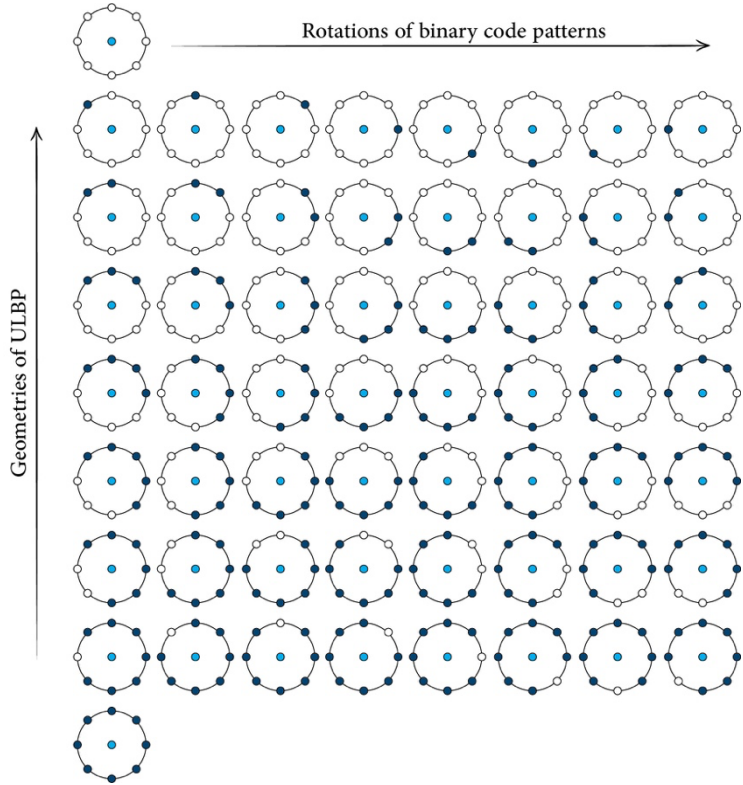

**Fig. 2.** Geometric representation of ULBP method. We can select one or more of these geometries to select landmarks from digital mammograms to construct PD. Different rows correspond to different types of texture. For example, the first and the last row correspond to flat and spot texture, row 5 (4 ones in the binary code) correspond to edges and row 6 correspond to corners.

For example, candidate landmarks are central pixel positions of the first rotation of $G_1(R_1)$ if a 3-by-3 patch's binary code is 00000001 where $R_\xi$ refers to rotations of specific $G_\lambda$ for $\xi = 1,2,...8$. We follow the same strategy to select a set of pixel value locations for each of the 56 ULBP geometries depicted in **Fig 2**.

The first two stages of **Fig. 3** show an example of a set of landmark pixel locations extracted from a normal and abnormal mammogram and their corresponding PDs. After selecting a set of mammogram pixel landmarks, we generate a Euclidean distance matrix $D$ from these pixel value locations which will then be used as input for PH generation pipeline.

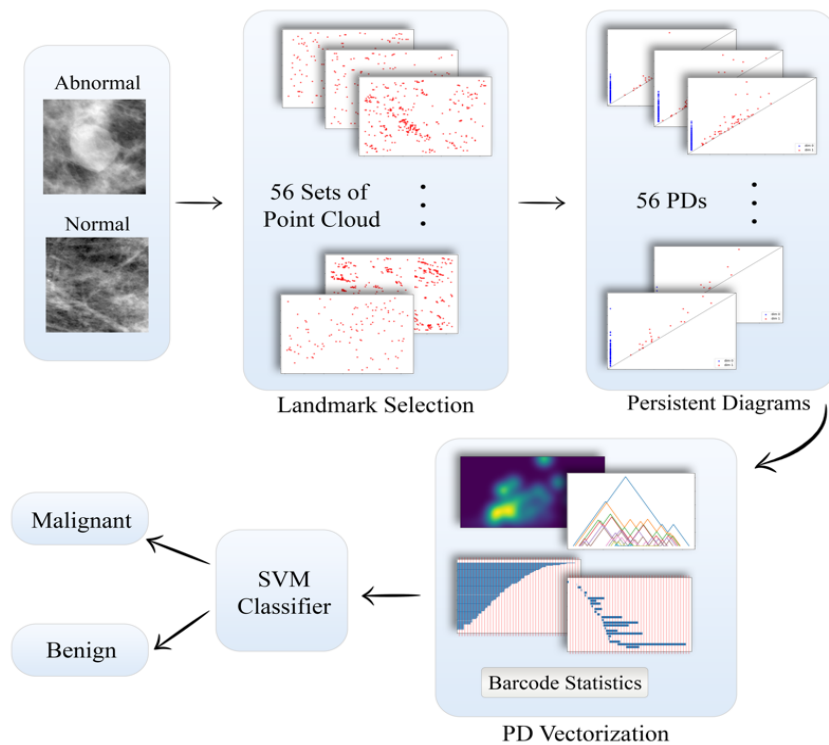

**Fig. 3.** Landmark based PH construction and classification pipeline.

### 2.2 Persistent Homology of Digital Images

In this section we introduce Vietoris-Rips simplicial complexes and cubical complexes as two persistent homology approaches to build topological features from breast mammograms.

### 2.2.1 Vietoris-Rips Complexes based on Image Pixel landmarks

In order to build topological features from data (point cloud or image), PH relies on mathematical objects known as simplices which are building blocks of higher dimensional objects in space known as simplicial complex. In this work we construct Vietoris-Rips ($VR$) simplicial complexes using the pixel-value locations obtained from ULBP method. For a set $\mathcal{L}$ of pixel landmarks in $\mathbb{R}^2$, its $VR$ with parameter $\epsilon$, denoted as $VR(\mathcal{L}, \epsilon)$, is the simplicial complex where $\{l_0, l_1, l_2, ..., l_\eta\}$ is its vertex set that spans a $\eta$-simplex iff the Euclidean distance between any two landmark locations is less than or equal to the chosen value of $\epsilon$, i.e $d(l_i, l_j) \leq \epsilon \ \forall \ 0 \leq i, j \leq \eta$. As we increase the value of $\epsilon$, so does the $VR$ of pixel locations. This process results in producing a nested sequence of $VR$ simplicial complexes known as filtration. In other words, $VR(\mathcal{L}, \epsilon_1) \subseteq VR(\mathcal{L}, \epsilon_2)$ if $\epsilon_1 \leq \epsilon_2$. Homological features born and vanished during filtration process are then stored as points in PD.

We direct interested reader to see [10], [11] , [1] for more mathematical details on $VR$ construction from a point cloud and PH introduction and mathematical backgrounds.

### 2.2.2 Cubical Complexes of Digital images

The cubical analogue of a (VR) simplicial complex is a cubical complex, in which the role of simplices is played by cubes of different dimensions, as in **Fig. 4**. A finite cubical complex in $R^d$ is a union of cubes aligned on the grid $\mathbb{Z}^d$ satisfying some conditions similar to the simplicial complex case [20]. A $d$-dimensional image is a map $\eta: I \subseteq \mathbb{Z}^d \to \mathbb{R}$. An element $v \in I$ is called voxel, or a pixel when $d = 2$, and $\eta(v)$ is called its greyscale value. There are several ways to represent digital images as cubical complexes, but greyscale image comes with a natural filtration and hence adopted here. Voxels are represented by vertices and cubes are built between those vertices. We represent voxels by $d$-cubes and all of its adjacent lower dimensional cubes are added. Next, we obtain a function on the resulting cubical complex $\mathbb{K}$ by extending the values of voxels to all the cubes $\sigma \in \mathbb{K}$ as follows:

$$\eta'(\sigma) := \min_{\sigma \ face \ of \ \tau} \eta(\tau)$$

Assume $\mathbb{K}$ to be the resulting cubical complex built on the greyscale image $I$. Let

$$\mathbb{K}_i := \{\sigma \in \mathbb{K} \ | \eta'(\sigma) \leq i \}$$

Be the $i$-th sublevel set of $\mathbb{K}$. The set $\{\mathbb{K}_i\}_{i \in Im(I)}$ defines a filtration of cubical complexes indexed by the value of greyscale function $\eta$.

### 2.3 Persistence Diagram Vectorization

Topological features summarized by PDs are not amenable to many machine learning and statistical tasks; for instance PD's Fréchet mean is not unique [12]. Hence, many vectorization approaches proposed to transform the data in PDs to resolve this issue and be able to apply machine learning methods. We use 4 methods to vectorise topological features in PD: Persistence Images [13], Persistence Landscapes [14], Binning [15] and Barcode Statistics. Next, we briefly describe each of these vectorization approaches.

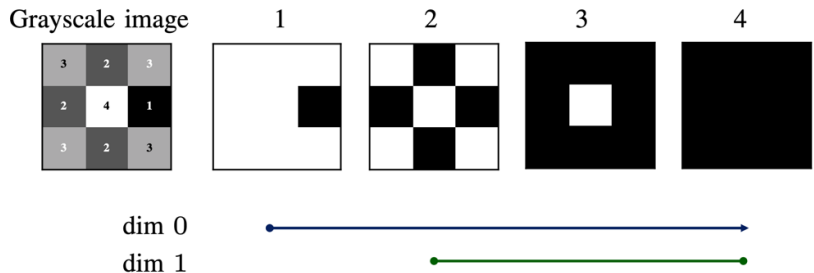

**Fig. 4.** A greyscale image patch and its corresponding cubical complex filtration and persistent barcode representation in dimension zero and one [20].

**Persistence Landscape (PL)** PL is one of the early vectorization methods proposed to map PDs into a stable and invertible function space using a family of piece wise linear functions $\{\Psi_k: \mathbb{R} \to \mathbb{R}\}_{k \in \mathbb{Z}}$ such that $\Psi_k(\tau) \coloneqq \sup\{m \geq 0 \mid \alpha^{\tau-m,\tau+m} \geq k\}$ where $\alpha^{i,j} \coloneqq \#\{P = (p_1, p_2) \in PD \mid p_1 \leq i \leq j \leq p_2\}$. More details of this method can be seen in [14]. Restricting these functions to a closed interval of $(a, b) \subset \mathbb{R}$ and choosing a uniform discretization will result in a 2-dimensional feature vector suitable for machine learning classifiers. In this paper, we set $k = 100$ to use the 100 largest such functions in our analysis of mammogram classification.

**Persistence Image (PI)** PI is one of the popular vectorization methods used to transform the topological information contained in PDs into a vector. To construct PI, first rotate PD by $\pi/4$ then turn the rotated PD into a persistent surface via $\Phi: \mathbb{R}^2 \to \mathbb{R}$ and a Gaussian distribution $\Phi_\mu$ such that $\Phi(PD) = \sum_{\mu \in PD} w(\mu) \, \Phi_\mu(z)$ where $w$ is a piecewise linear weight function. Finally, discretize the persistent surface $\Phi$ by taking samples over a regular grid.

**Persistence Binning (P-binning)** This approach is one of the simple vectorization methods that relies on counting the number of bars in PBs that intersects with each vertical line $V = 0,1,2,\ldots,\omega$. In this paper we set $\omega = 30$ equidistance vertical lines. Thus, a topological feature vector of size ω obtained for different dimensions of PBs.

**Barcode Statistics (P-statistics)** The simplest approach to vectorise the space of PBs is to extract statistics directly from PBs. We collect only 10 statistics: average and standard deviation of birth, death and lifespan of bars, median of birth, death and lifespan of bars and finally the number of bars. Statistics of birth of topological features in dimension zero of PBs can be ignored as they return zero by default.

## 3    Dataset Description and Evaluation scheme

Two widely used mammogram databases utilised to test the performance of landmark-based PH for mammogram abnormality classification which are publicly available. The two datasets are known as Digital Database for Screening Mammogram (DDSM) [16] and Mini Mammographic Image Analysis Society (Mini-MIAS) [17]. Mini-MIAS dataset contains 113 abnormal and 209 normal mammograms of women breast which include fatty, granular, calcification, architectural symmetry and dense cases. DDSM constitutes of 2620 mammograms in total in which 512 mammograms randomly selected in our experiments with 302 normal cases and 257 abnormal cases. Images in both datasets are cropped region of interest (ROI) images of size 128-by-128. A number of benchmarking mammographic datasets are available for experimental purpose in which they vary according to certain pre-defined criteria like type and structure of the digital mammogram, dense, fatty or glandular tissues, noise level in the images and the number of benign and malignant cases in these datasets. We opt to use Mini-MIAS and DDSM due to the fact that images in both datasets captured in uncontrolled conditions, images contain sufficient noise and low-resolution images. Examples of images from Mini-MIAS and DDSM datasets can be seen in **Fig. 5** and **Fig. 6**.

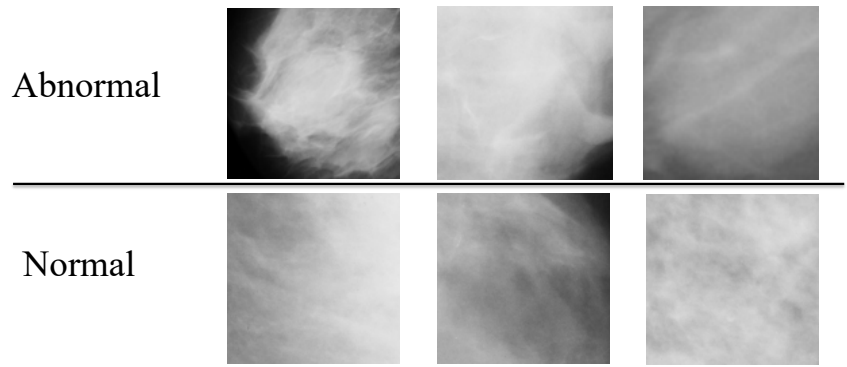

**Fig. 5.** Examples of ROI for Normal and Abnormal cases from Mini-MIAS dataset.

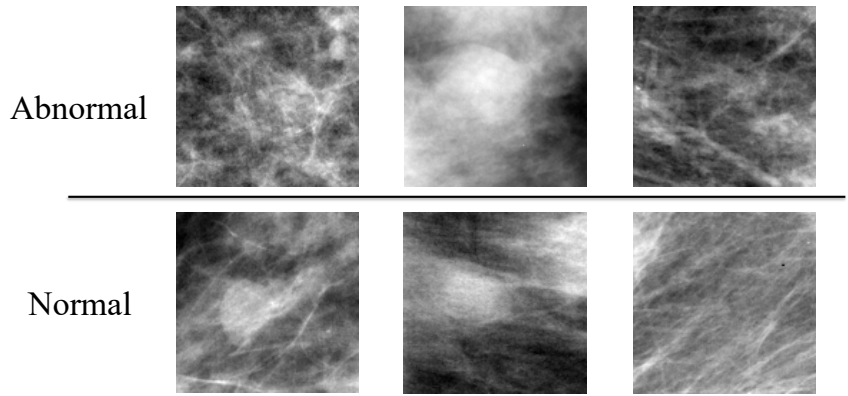

**Fig. 6** Examples of ROI for Normal and Abnormal cases from DDSM dataset.

Two evaluation metrics used which are sensitivity (SE), the proportion of breast cancer cases correctly classified as patients having malignant tumors, and specificity (SP) which corresponds to the number of normal breast mammogram cases correctly classified as normal.

The formula for both sensitivity and specificity is defined as follows:

$$Sensitivity = \frac{True\ positive}{True\ positive + False\ negative}$$

where True Positive (TP) refers to cancer patients truly identified as patients having abnormal breast mammograms, False Negative (FN) is the breast cancer patients misclassified as negative of having breast cancers.

$$Specificity = \frac{True\ negative}{True\ negative + False\ positive}$$

where true negative (TN) refers to number of truly classified women clear of the breast cancer and false positive (FP) means the number of cases wrongly classified as breast cancer positive which in fact are clear of having cancer.
Support vector machine (SVM) classifier used to differentiate abnormal mammograms from their normal counterparts where we optimized all hyperparameters of SVM with a 5-fold cross validation setting.

## 4      Reproducibility and Implementation Details

In all experiments we extracted 0-dimensional and 1-dimensional PDs using Ripser package in python (https://pypi.org/project/ripser/). PI of resolution 30-by-30, linear weighting function and the rest of the parameters in default setting generated using GUDHI library in python (https://pypi.org/project/gudhi/). PL generated with $k = 100$ and the rest of other parameters with default setting from GUDHI library. Cubical complex filtration and its corresponding PD constructed using GUDHI library in python. SVM classification performed in MATLAB with standardization and tuning for optimal hyperparameters. In other words, in each fold we search for the best kernel among the four kernel options available in MATLAB which are linear, gaussian, radial basis function and polynomial. This means that a linear kernel for the 1[st] fold may not be good in the 2[nd] fold and we may have a case of 4 different kernels in a 5-fold cross validation. ULBP landmarks selected from breast mammograms using DAAR software [18]. Padding of zero is performed during the process of ULBP landmark selection during 3-by-3 patch scanning of mammograms with overlap value of 2 between two consecutive patches.

## 5      Experimental Results

In order to classify mammogram scans, SVM classifier trained in a 5-fold cross validation setting based on topological features. For each image, there would be 56 PDs build on 56 point-clouds extracted from ULBP landmark selection method. There are many approaches one can train and test machine learning classifier for the 56 vectorized PDs generated. We first concatenated the topological features following the 7 geometrical groups in ULBP. In other words, features of the 8 rotations of each ULBP geometry concatenated for each of PD vectorization method. In addition to the 7 feature vectors obtained, PH features extracted in dimension zero and one which correspond to connected components and 1-dimensional cycles. Experimental results obtained from combining topological features in dimension zero and one was better than using either dimensions alone. In **Table 1** and **Table 2**, we report sensitivity and specificity of best performing ULBP geometry obtained from best performing dimension of PH features and the 4 PD vectorization methods. Out of the 7 ULBP geometries, none of the geometries performance is consistence on both datasets using either of the 4 vectorization methods utilised. PL with G3 performed better than other ULBP geometries on DDSM while P-Binning and G7 performed best on Mini-MIAS dataset.

**Table 1.** Top performing ULBP geometries and PH dimension and all PD vectorizations for DDSM. Avg = average, Std = standard deviation for 5-folds cross-validation using SVM.

| Feature type | Classification Metrics | Avg ± Std |
|---|---|---|
| PD-dim0,1 & P-Binning & $G_5$ | Sensitivity | 85.02 ± 7.5 |
|  | Specificity | 77.4 ± 2.7 |
| PD-dim1 & P-Statistics & $G_3$ | Sensitivity | 85.1 ± 4.9 |
|  | Specificity | 79.7 ± 6.6 |
| PD-dim0,1 & PI & $G_3$ | Sensitivity | 76.4 ± 9.4 |
|  | Specificity | 66.9 ± 7.3 |
| PD-dim0,1 & PL & $G_3$ | Sensitivity | **86.06 ± 4.8** |
|  | Specificity | **80.9 ± 4.4** |

**Table 2.** Top performing ULBP geometries and all PD vectorizations for Mini-MIAS.

| Feature type | Classification Metrics | Avg ± Std |
|---|---|---|
| PD -dim0 & P-Binning & $G_7$ | Sensitivity | **97.6 ± 1.5** |
|  | Specificity | **95.5 ± 3.2** |
| PD -dim0,1 & P-Statistics & $G_5$ | Sensitivity | 98.6 ± 2.9 |
|  | Specificity | 94.6 ± 3.8 |
| PD -dim0,1 & PI & $G_7$ | Sensitivity | 98.1 ± 1.0 |
|  | Specificity | 94.6 ± 2.1 |
| PD -dim0,1 & PL & $G_7$ | Sensitivity | 97.6 ± 0.1 |
|  | Specificity | 92.8 ± 6.1 |

Combined PH features of dimension zero and one for all 56 ULBP geometries with PL provided 92% sensitivity and 86% specificity for DDSM, see **Table 3**. The results reported here can be partially compared with that reported in [19] where ULBP and PH used to mammogram abnormality classification.

**Table 3.** Concatenation of all ULBP geometries together with dimension 0 and 1 of PD for DDSM classification using top 3 PD vectorization.

| Feature type | Classification Metrics | Avg ± Std |
|---|---|---|
| PD-dim0,1 & PL | Sensitivity | **92.3 ± 4** |
|  | Specificity | **86.5 ± 3** |
| PD-dim0,1 & P-Binning | Sensitivity | 85.1 ± 6 |
|  | Specificity | 82.6 ± 4 |
| PD-dim0,1 & P-Statistics | Sensitivity | 87.6 ± 4 |
|  | Specificity | 82.3 ± 4 |

Authors in [19] only used binning to vectorize PD with KNN classifier and they reported best classification performance of a sensitivity of 86% and specificity of 98% for Mini-MIAS together with 82% sensitivity and 75% specificity for DDSM. Our results outperform these results in both datasets.
Finally, in **Table 4** and **5**, we report the classification performance of SVM using cubical complexes filtration approach where we used all grayscale pixel values of the mammograms to construct one PD and then the 4 vectorization methods.

**Table 4.** Cubical Complex performance results for **Mini-MIAS** dataset using 4 different vectorization methods and 3 homology dimensions.

| Feature Type | Sensitivity (Avg ± STD) | Specificity (Avg ± STD) |
|---|---|---|
| P-Binning and PD-dim0 | 99.02 ± 2.18 | 2.51 ± 3.65 |
| P-Binning and PD-dim-1 | 99.02 ± 1.34 | 0.91 ± 2.03 |
| P-Binning and PD-dim0,1 | 98.54 ± 2.18 | 2.51 ± 3.65 |
| P-Statistics and PD-dim0 | 98.54 ± 1.34 | 92.15 ± 3.25 |
| P-Statistics and PD-dim1 | 98.09 ± 1.07 | 96.47 ± 1.99 |
| P-Statistics and PD-dim0,1 | **98.58 ± 1.3** | **94.76 ± 1.54** |
| PI and PD-dim0 | 88.68 ± 5.53 | 78.04 ± 13.48 |
| PI and PD-dim1 | 90.95 ± 4.55 | 84.98 ± 3.81 |
| PI and PD-dim0,1 | 93.83 ± 2.53 | 91.24 ± 2.67 |
| PL and PD-dim0 | 93.39 ± 3.23 | 69.96 ± 12.29 |
| PL and PD-dim1 | 89.7 ± 8.78 | 78.51 ± 8.56 |
| PL and PD-dim0,1 | 95.43 ± 6.23 | 80.22 ± 11.68 |

**Table 5.** Cubical Complex performance results for **DDSM** dataset using 4 different vectorization methods and 3 homology dimensions.

| Feature type | Sensitivity (Avg ± STD) | Specificity (Avg ± STD) |
|---|---|---|
| P-Binning and PD-dim0 | 57.08 ± 15.13 | 62.94 ± 10 |
| P-Binning and PD-dim-1 | 68.6 ± 17.5 | 62.26 ± 12.89 |
| P-Binning and PD-dim0,1 | 69.16 ± 19.52 | 51.14 ± 22.9 |
| P-Statistics and PD-dim0 | 82.08 ± 7.57 | 80.57 ± 5.58 |
| P-Statistics and PD-dim1 | 80.13 ± 8.5 | 87.19 ± 3.88 |
| P-Statistics and PD-dim0,1 | **86.03 ± 8** | **81.72 ± 4.89** |
| PI and PD-dim0 | 61.01 ± 9 | 81.69 ± 7.85 |
| PI and PD-dim1 | 71.62 ± 10.01 | 78.19 ± 5.35 |
| PI and PD-dim0,1 | 73.19 ± 6.18 | 73.95 ± 6.48 |
| PL and PD-dim0 | 74.46 ± 7.17 | 71.54 ± 7.86 |
| PL and PD-dim1 | 81.13 ± 6.06 | 73.53 ± 7.34 |
| PL and PD-dim0,1 | 83.74 ± 4.72 | 77.01 ± 5.03 |

It can be seen that using cubical complexes we can obtain 98% and 94% for sensitivity and specificity, respectively for Mini-MIAS dataset and up to 86% of sensitivity and 81% specificity for DDSM. P-Statistics performed better than the other 3 vectorization methods using cubical complexes which was not the case using landmark-based VR filtration. Using our proposed landmark based approach, one geometry alone ($G_7$) achieved roughly the same performance for Mini-MIAS where only use a small portion of the mammogram scan pixel values. For DDSM, we outperform cubical complexes if we concatenate topological features from all geometries and obtain almost the same performance using one geometry, i.e. $G_3$.

## 6    Discussion and Future work

This study introduced a distributed method of constructing 56 PDs based on automatically extracted landmarks from breast mammograms. In general, we have found that a small set of pixel landmarks is enough to detect abnormality in breast mammograms such as $G_3$ in DDSM and $G_7$ in Mini-MIAS dataset. Computing the 56 PDs can be done in a distributed manner which is crucial for large scale datasets. Instead of building one PD using cubical complexes, our approach provides a localized PD representation that convey topological features linked to different types of mammogram texture distribution. Different PD vectorizations examined where we found that it is a good practice to try more than one method as a single approach may not consistently perform well on different datasets. This work is the first step towards a more comprehensive study for different approaches of PD vectorization in medical imaging because we concluded that different types of vectorization methods affect the performance greatly as it can be seen from **Table 1** to **Table 5**. Until now, to the best of our knowledge, there is no comprehensive analysis or a roadmap to select suitable vectorization method(s) for medical image analysis or any other image modalities. On the other hand, it is not an easy task to search the rich literature of PH vectorizations and pick the correct method suitable for the problem in hand. Nonetheless, based on the findings in this work, at this stage we are not advising to use a single vectorization method as this could lead to misleading performances.

Furthermore, our analysis showed that proposed landmark based PH can outperform classical approaches of building topology from digital images such as cubical complexes. This points to the fact that a small set of pixel value landmarks that correspond to different type of textures can be used to differentiate malignant mammogram scans from benign scans. This is particularly useful when the medical image dimensions (number of rows and columns) are very high and using the entire pixel values is time consuming or down sampling it may result in loss of critical medical information.

The only limitation of this work is the increase in dimension of feature vectors when combining more than one ULBP geometry as it was the case when we concatenated all ULBP geometries to boost the classification performance for DDSM dataset in **Table 3**. Aggregating PDs before vectorization is one approach to address this limitation in future. Future work will also focus on using other texture methods as landmark selection procedure such as center-symmetric LBP or small image patches of high density. Furthermore, testing proposed ULBP based PH on other medical image modalities such as ultrasounds and other types of disease is included in our list of future works.